\documentclass[12pt,preprint]{aastex}

\received{2003 August 21}
\begin{document}
\slugcomment{Submitted to ApJ part 1}
\title{The optical counterpart of an Ultra-luminous X-Ray Source in NGC 5204}

\author{Ji-Feng Liu, Joel N. Bregman and Patrick Seitzer}
\affil{Astronomy Department, University of Michigan, MI 48109}

\begin{abstract}

Ultra-luminous X-Ray sources are extra-nuclear point sources in external
galaxies with $L_X=10^{39}$--$10^{41}$ erg/s  and are among the most poorly
understood  X-ray sources.  To help understand their nature, we are trying to
identify their optical counterparts by combining images from the Hubble Space
Telescope and the Chandra Observatory. Here we
report upon the optical counterpart for the ULX in NGC 5204, which has average
X-ray luminosity of $\sim3\times10^{39}$ erg/s and has varied by a factor of
50\% over the last 10 years. A unique optical counterpart to this ULX is found
by carefully comparing the Chandra ACIS images and HST WFPC2 and ACS/HRC
images. The spectral energy distribution and the HST/STIS FUV spectrum of this
object show that it is a B0 Ib supergiant star with peculiarities, including
the $\lambda$1240 N V emission line that is uncommon in B stellar spectra but
has been predicted for X-ray illuminated accretion disks and seen in some X-ray
binaries.  Study of its FUV spectrum leads to a binary model for this ULX in
which the B0 Ib supergiant is overflowing its Roche Lobe and accreting onto the
compact primary, probably a black hole.  This picture predicts an orbital
period of $\sim10$ days for different black hole mass, which can be tested by
future observations.

\end{abstract}


\section{INTRODUCTION}

X-ray observations have revealed two categories of accretion powered X-ray
point sources: the stellar mass X-ray binaries and the X-ray emission from
active galactic nuclei. The first class has a typical luminosity range of
$10^{33}$--$10^{38}$ erg/sec, powered by accretion onto a compact object such
as a white dwarf, a neutron star or a black hole in a binary system. The second
class usually has an X-ray luminosity of $>10^{41}$ erg/sec, powered by
accretion onto a central super-massive black hole of mass $10^6$--$10^9$
$M_\odot$.  X-ray point sources with luminosities within the range of
$10^{39}$--$10^{41}$ have also been observed in some external galaxies, first
by the Einstein X-ray satellite (cf. Fabbiano 1989), later on by the ROSAT and
ASCA X-ray satellites (e.g., Colbert \& Mushotzky, 1999; Roberts \& Warwick,
2000; Mizuno et al., 1999), and now by the Chandra X-ray satellite (e.g., Zezas
et al., 2002). Some of these objects are nuclear sources and may be low
luminosity AGNs, but others are extra-nuclear sources and form a different
class. This class of extra-nuclear luminous sources has several names, such as
intermediate luminosity X-ray objects (IXOs) and Ultra-luminous X-ray sources,
or ULXs, which we use in this paper.

The observed luminosities of ULXs, if the X-ray emission is isotropic, require
accreting black holes of masses $>10^3$ $M_\odot$ if emitting at the level of
$10^{-2}$ of the Eddington luminosity. Such black holes, if they exist, are the
missing links between stellar mass black holes and super-massive black holes in
the nuclei of galaxies. This idea is consistent with the X-ray spectral
analyses of some ULXs in nearby galaxies (e.g., Makishima et al.  2000).
However, the formation of such massive black holes is not predicted by stellar
evolution theory, and it is still in debate about the prospect to form such
objects in dense star clusters. Alternatively, these sources may be stellar
mass black holes or neutron stars whose emission is beamed mildly or
relativistically, thereby representing micro-quasars (King et al. 2001;
Georganopoulos et al. 2002). If the emission is beamed, the intrinsic
luminosities become sub-Eddington, and there are known examples of beamed
Galactic X-ray sources (c.f., Mirabel \& Rodriguez 1999). It is also possible
that the luminosities are truly super-Eddington, for example, obtainable from
accretion disks with radiation-driven inhomogeneities (Begelman, 2002).

To identify the ULXs at other wavelengths such as optical will supply
information that is helpful to uncover the nature of these objects.  For
example, Pakull \& Mirioni (2002) observed a sample of ULXs with ground-based
telescopes, and were able to associate a number of ULXs to emission nebulae
with kinematic ages of some million years. For the ULX positions, they used
those derived from ROSAT HRI observations, which usually have errors of
$\sim5^{\prime\prime}$. Such a position error corresponds to 250 pc at a
distance of 10 Mpc, and makes it impossible to identify down to single sources.
However, with the advent of Chandra X-ray satellite, identifications down to
single sources become possible owing to the sub-arcsecond spatial resolution of
Chandra observations.  Some efforts have already made along this line of
thoughts. For example, by comparing Chandra ACIS-S and HST/WFPC2 observations
of the ''Antennae'' starburst galaxies, Zezas et al.  (2002) found that 10 out
of 14 ULXs are associated with stellar clusters, with offsets between the X-ray
sources and the clusters less than $2^{\prime\prime}$.  Wu et al.  (2002) found
that the ULX in edge-on spiral galaxy NGC4565 is associated with a stellar
cluster on the outskirt of the bulge. Also, as part of our search for optical
counterparts of ULXs in nearby galaxies, we (Liu, Bregman \& Seitzer, 2002)
found for the ULX in M81 a unique optical counterpart that is an O8 star, which
is probably accreting mass onto a black hole of 20 $M_\odot$  via Roche lobe
overflow.

In this paper we report on the optical counterpart for the ULX in NGC 5204.  
In the last 10 years this ULX was observed twice by ROSAT PSPC, twice by ROSAT
HRI, and twice by Chandra ACIS, with an average $L_X \sim 3\times10^{39}$ erg/s
and a variation of 50\%. 
With positions from Chandra ACIS observations, Roberts et al. (2001) and Goad
et al. (2002) identified this ULX with three adjacent objects on HST WFPC2
images, the brightest one of which they estimated to be a young cluster with a
few thousand member stars. 
Our re-analysis of the same data sets show that the most likely optical
counterpart should be the faintest of the three objects, which is confirmed by
new HST/STIS and HST ACS/HRC data as a single B0 Ib star. 
In $\S$2, we present the observations utilized, the data analysis procedures,
and the results. We discuss the implications of its optical identification in
$\S$3. For the distance to NGC 5204, we use 4.3 Mpc (Tully 1992).  At such a
distance, 1 HRC pixel, i.e., $0\farcs027$ which provides critical sampling at
6300 \AA, corresponds to a physical scale of 0.6 parsecs.

\section{OBSERVATIONS AND DATA ANALYSIS}

\subsection{Data processing}

A central issue in comparing optical and X-ray imaging data is registering the
X-ray positions on the optical images, which requires accurate astrometry for
both images. Here we use Chandra observations for X-ray images and HST
observations for optical images, because these two instruments have the best
astrometric accuracy available.

Two archive Chandra ACIS observations (Observation ID 2028, observed on 2001
January 9; Observation ID 2029, observed on 2001 May 2) were used to obtain the
X-ray position of the ULX.  
To detect all discrete sources, we employed the CIAO tool {\tt Wavdetect} on
ACIS-S chips, although any detection method would find nearly all the same
sources.  
Detected in both observations are the ULX and a second source
$\sim20^{\prime\prime}$ away, which is designated as CXOU J132940.1+582454
following the Chandra source naming convention. Both sources were reported in
Roberts et al. (2001).
The average X-ray position of the ULX is R.A. = 13:29.38.62, Decl = 58:25:5.6;
the positions in two observations differ by $0\farcs4$, within the typical
R.M.S.  of absolute ACIS positions ($0\farcs6$; Aldcroft et al. 2000).

NGC 5204 was observed with HST twice, the first time in May, 2001 by HST WFPC2
using filter F606W and F814W(I),  as part of a different program.  
Cosmic rays were not removed from the images since the observations were single
exposures.  Both the ULX and J132940.1+582454 fall on the WF2 chips of these
WFPC2 observations.  
Iraf tasks {\tt metric/invmetric}, which correct the geometric distortion
introduced by the camera optics, were used to transform between WFPC2 image
positions (X,Y) and celestial coordinates (RA,DEC).  
The final relative positions are accurate to better than $0\farcs005$ for
targets contained on one chip, and $0\farcs1$ for targets on different chips.
The absolute accuracy of the positions obtained from WFPC2 images is typically
$0\farcs5$ R.M.S. in each coordinate (HST data handbook).

NGC5204 was also observed with the ACS/HRC on board HST in November 2002 for
our program.  
The HRC image has a better spatial resolution, with a pixel size of
$0\farcs027$ that samples the HST point spread function critically at 6300 \AA.
HRC images were obtained in two filters, F220W and F435W, and for the F220W
image, we obtained several images in a small raster pattern to better sample
the PSF.  
The exposures were processed with multidrizzle\footnote{see
http://stsdas.stsci.edu/pydrizzle/multidrizzle/} to combine the dithered images
and remove the cosmic rays. 
Both the ULX and J132940.1+582454 lie in the HRC chip, leading to an accurate
relative astrometric solution.

Direct comparisons between ACIS images and HST WF2 and ACS/HRC images yield
several candidates of optical counterpart to the ULX.  
Registering the nominal X-ray position of the ULX onto the WF2 imageis (Figure
1), one finds within the $1^{\prime\prime}$ nominal error circle three objects,
which were designated as HST-1, HST-2, and HST-3 by Goad et al. (2002). 
For HRC images (Figure 1), the nominal error circle includes HST-3 (which is
designated as U1), while HST-1 and HST-2 (which breaks down to a chain of
separate sources) are outside of this error circle. This indicates a pointing
offset of $\sim1\farcs5$ between WFPC2 and ACS observations.

To find a unique optical counterpart to the ULX, we improved the positional
accuracy further by relative astrometry between the ULX and J132940.1+582454,
which fall on the same ACIS-S chip, the same WF2 chip, and the same ACS/HRC
chip. 
Inspections of the WF2 images show that there is a source falling within the
$1^{\prime\prime}$ error circle of the nominal J132940.1+582454 position
(Figure 2).  
On the HRC images, this same source falls within the $0\farcs6$ error circle of
the nominal J132940.1+582454 position. 
This source is a point source on HRC images; its non-pointlike appearance in
the F606W image may be due to cosmic rays. This source is blue, with magnitudes
listed in Table 2, and has a similar color as U1. Given its magnitude, it can
not be an O/B star within our Galaxy, but it can be an O/B star in a high mass
X-ray binary (HMXB) system in NGC 5204, or a background AGN. In both cases, it
could be a strong X-ray emitter.
We believe it is the optical counterpart to J132940.1+582454, considering the
low surface density of objects with similar brightness or brighter in its
vicinity, which leads to a small chance (less than 1\%) for it to be a random
object falling into the error circle of J132940.1+582454.  
Given the WF2 position and HRC position of this optical counterpart to
J132940.1+582454, and the X-ray position of J132940.1+582454, the ULX positions
on WF2 images and HRC images were derived from its X-ray position (Table 1). 
The position error mainly comes from the centroiding error of J132940.1+582454.
For the separation between the ULX and J132940.1+582454, the ACIS plate scale
variation produces an error of $0\farcs02$.  The rotation R.M.S.  for both
WFPC2 and ACIS-S chips are $\sim0\fdg01$, and the error introduced by this
effect is negligible. The final position error is $\Delta\alpha = 0\farcs25,
\Delta\delta = 0\farcs16$.

\subsection{Analysis of the Images}

The much reduced error circles were overlayed on the WF2 images and HRC images
in search of the optical counterpart to the ULX (Figure 1). 
On the WFPC2 images, while there are three objects within the
$1^{\prime\prime}$ nominal error circle, among which HST-1 is the brightest,
HST-3 is the only object (partly) within the reduced error circle, suggesting
HST-3 is the unique optical counterpart to the ULX. 
On the HRC images, at the corresponding position of HST-3 within the reduced
error circle is a point-like source (U1). On the edge of the error circle is
another source, designated as U2, which is much fainter than U1, and marginally
discernible in the WF2 images.
HST-1 is $\sim1\farcs5$ away from the reduced error circle; it does not even
fall within the $1^{\prime\prime}$ nominal error circle like it does in the WF2
images.

The optical counterpart U1 is a point source instead of an extended source. 
On the F606W WF2 image, there is a spurious tail to the south-east of HST-3;
however, this feature does not appear in the F814W WF2 image or the HRC images,
and probably is a cosmetic feature due to cosmic rays.
To study its spatial extent, we constructed the radial profiles for U1 as in
the F220W and F435W HRC images (Figure 3). Its profiles are compared to the
radial profiles for a nearby extended source, presumably a star cluster, and to
a collection of 23 point-like sources in the proximity of the ULX. 
For the F220W HRC image, point-like objects have an average full-width at half
measure (FWHM) of 3.40$\pm$0.24 pixels, while U1 has a FWHM of 3.55 pixels, and
the cluster has a FWHM of 5.68 pixels; for the F435W HRC image, point-like
objects have an average FWHM of 2.47$\pm$0.24 pixels, while U1 has a FWHM of
2.40 pixels, and the cluster has a FWHM of 4.02 pixels.  On both images, the
optical counterpart U1 is consistent with being a point source, and can be
distinguished easily from an extended source.

HST-1 is a composite source as seen from the HRC images.
Based on the HST WFPC2 observations, Goad et al. (2002) concluded that HST-1 is
a young cluster composed of a few thousand stars.  
However, in the HRC images with higher spatial resolution, HST-1 can be
decomposed into two components, one bluer component partly overlapping another
redder component whose center is about 7 pixels (i.e., 4.2 parsecs) away. Such
a decomposition is clearly shown in the contour plot (Figure 4) of the region
around HST-1 and U1, which also shows that while the redder component is
point-like, the bluer component shows complex infrastructures and is certainly
not a single point source.

We can construct a spectral energy distribution (SED) from the four wide-band
photometry measurements, i.e., ACS/HRC F220W (NUV), F435W, and WFPC2/WF2 F606W
and F814W. 
For the ACS/HRC observations, aperture photometry was carried out with a 5
pixel annular aperture,  and a 5\% nominal error is adopted. For the WFPC2
observations, HSTphot (Dolphin 2000) was used to perform PSF fitting photometry
to extract photon counts that are converted to STMAG.  
The results are listed in Table 2 for U1, U2, and HST-1, and  the curves of the
fluxes for the three objects are plotted in Figure 5.  
These curves show a trend of secular decline, reflecting the blackbody
nature of the radiation. Note that the F606W and F814W fluxes, obtained from
WF2 images, may contain contributions of nearby faint sources that can only be
resolved in HRC images.
The curve for HST-1 show a significant bump at redder wavelength as compared to
the Rayleigh-Jeans extension of its NUV light. These extra light can be
attributed to its redder component, while the NUV light can be mostly
attributed to its bluer component.

The flux curves are compared to those of standard stars to shed light on the
spectral types of these objects. 
These standard stars, whose absolute magnitudes and colors in the
Johnson-Cousins system are taken from Schmidt-Kaler (1982), are placed at the
distance to NGC 5204, and corrected for extinction toward U1 of $n_H = 10^{21}$
cm$^{-2}$ assuming Galactic relations $n_H = 5.8\times10^{21}E(B-V)$ (Bohlin et
al. 1978) and $R_V=3.1$. The adopted $n_H$ value is inferred from the X-ray
spectrum of the ULX, which can be fitted by an absorbed power law model with
$n_H = 1.6\pm0.3\times10^{21}$ cm$^{-2}$ or an absorbed thermal bremsstrahlung
model with $n_H = 5.4\pm1.7\times10^{20}$ cm$^{-2}$ (Roberts et al. 2001).
The magnitudes and colors for U1 from the HRC and WFPC2 observations are
consistent with those of stars of type O5 V, O7 III, or B0 Ib. The degeneracy
of these different possibilities is lifted when the FUV spectrum of U1 is
considered (see below).
The fluxes from HST-1 are a few times brighter than U1 in the F220W band,
making it impossible to be explained by a single O/B star.  Combining with its
complex non-pointlike infrastructures in the F220W band, we conclude the blue
component of HST-1 is an OB association of at least a few O/B stars. The fluxes
in the F606W and F814W bands, mostly from the red component, are more than 10
times higher than U1, and can be attributed to a young cluster of a few
thousand stars (Goad et al.  2002).

\subsection{Spectral Analysis}

To distinguish between O/B stars of different spectral types, MAMA/FUV spectral
observations were taken by HST STIS in November, 2002 for our program.  
The placement of the $50^{\prime\prime}\times0\farcs2$ slit used in the
observation is shown on the HRC images in Figure 1. 
The spectra of U1 and HST-1 were extracted and analyzed following standard
procedures, and plotted in Figure 6.
Galactic absorption lines such as $\lambda1260$ and $\lambda1527$ Si II lines
can be seen in both spectra. These lines are usually rather narrow due to the
low temperature of the absorbing medium;  stellar lines are usually from high
ionization levels and wider for the high stellar temperature.

%
As demonstrated in Figure 5, the U1 spectrum shows a spectral slope and flux
level that are consistent with the wide-band photometry measurements.
The STIS FUV spectrum of U1 was compared to the IUE stellar spectral
library\footnote{http://www-int.stsci.edu/~jinger/iue.html}. The U1 spectrum
declines steadily toward longer wavelength, and resembles the spectra of O
stars and B stars earlier than B8.  It is distinctly different from spectra of
B8 and later stars which decline toward shorter wavelength.  

More information can be obtained by comparing the spectral lines.
The strongest line feature in the spectrum of U1 is the $\lambda1299$ Si III
line, which is commonly seen in hot O/B stars, confirming its identification as
an O/B star. The equivalent width for this line is $\sim1.5$\AA, which is
suggestive of an O/B star cooler than about 25000 K (Prinja, 1990), so the most
likely identification of U1 is a B0 Ib star rather than an O5 V star or an O7
III star.
A low resolution model spectrum for a B0 I star from the Kurucz 1993
models\footnote{ftp://ftp.stsci.edu/cdbs/cdbs2/grid/k93models/} is over-plotted
in Figure 6 for comparison.
Other line features implying an O/B star include the $\lambda1175$ and
$\lambda1247$ C III lines, although these lines cannot be used to estimate the
stellar temperature due to the high noise level.

The U1 spectrum shows some peculiarities for a B0 Ib star.
The UV resonance lines of C IV $\lambda1550$ and SI IV $\lambda1400$, with P
Cygni-type profiles indicative of high velocity hot O/B stellar winds, are
commonly seen in B0 Ib stars and other O/B stars in the stellar spectral
library.  However, these conspicuous features are missing from the U1 spectrum.
The $\lambda1306$ emission line complex, a blend of O I and Si III lines
commonly seen in optically thin HII regions, is present in the U1 spectrum.
However, this emission feature is usually not obvious in the spectra of O/B
stars in the spectral library. It may be because these standard stars are all
nearby Galactic stars, and a spectral observation includes only a small
fraction of the H II regions surrounding the O/B stars.
Another unusual feature is the high ionization $\lambda1240$ N V emission line,
which has an ionization energy (77.5 eV) above that of He II (54 eV). There is
very little such ionizing radiation from a B0 star, and this line is not seen
in normal B star spectra.  The species N V can be achieved by collisional
ionization at temperature  $\sim 2\times10^5$ K, which does not occur in a B0
Ib star.  
One place to achieve such high ionization states is in an accretion disk and
its corona, where the temperature can be higher than $10^7$ K. Rather strong
$\lambda1240$ N V emission is predicted for an X-ray illuminated accretion disk
and corona (Raymond, 1993), and it has been observed in low mass X-ray binaries
(LMXBs) such as LMC X-3 (e.g., Cowley et al. 1994) and Hercules X-1 (e.g.,
Boroson et al. 2001).

In contrast, the HST-1 spectrum shows all the usual features of O/B stars,
without the peculiarities. For example, the peculiar line features of
$\lambda1240$ N V and the $\lambda1306$ emission line complex present in the U1
spectrum do not show up in the HST-1 spectrum.  The $\lambda1550$ C IV and the
$\lambda1400$ Si IV absorption line features typical of O/B stars are clearly
present, revealing  O/B stars in this composite source. This is consistent with
our conclusion of HST-1 being an OB association plus a young cluster.

\section{DISCUSSION}

With improved astrometry, we found by combining Chandra ACIS data and HST WFPC2
and ACS/HRC data that the ULX in NGC 5204 has a unique optical counterpart U1.
The spectral energy distribution of U1 shows it is an O5 V, O7 III or a B0 Ib
star. 
With the help of its FUV spectrum, we found  that U1 is a B0 Ib star with some
peculiarities. For example, the resonance lines C IV $\lambda1550$ and Si IV
$\lambda1400$ commonly seen in O/B stars due to hot O/B stellar winds are
missing, while the $\lambda1306$ emission line complex commonly seen in
optically thin H II regions is present; the $\lambda1240$ N V emission line is
also present which indicates a temperature $\sim 2\times10^5$ K.
All these observations can be explained by a binary model where the
secondary of a B0 Ib supergiant is supplying fuel to the primary by Roche Lobe
overflow.
The observation of such a system supports the idea of King et al. (2001) that
ULXs can be HMXBs with evolved secondary stars that overflow their Roche lobes.

The observed spectral features can easily fit into the proposed binary
scenario.
Due to the Roche potential, matter escaping the surface of the B0 Ib star
flows rather smoothly along the equi-potential contours of the Roche Lobe, and
accrete onto the compact primary to form a disk. 
In the presence of an accretion disk and the large X-ray luminosity observed,
Nitrogen in the disk/corona can be easily photoionized and/or collisionally
ionized to N V, to give the  N V $\lambda1240$ emission line as observed in the
spectrum of U1.
The bulk of the matter does not escape to infinity, thus no large scale high
velocity stellar wind is formed as in the cases of usual O/B stars.  While
there is a small amount of  C IV $\lambda1550$ and Si IV $\lambda1400$ emission
from the photosphere in the non local thermal equilibrium state, it is
understandable that no significant such lines with P Cygni-type profiles emerge
in the spectrum of U1 in the absence of large scale stellar winds. Another
disadvantage against these lines is the presence of copious X-ray photons from
the accretion disk, which may ionize Carbon and Silicon to ions higher than C
IV and Si IV.
The H II region around the binary system,  with a size of $\sim10$ parsecs at
the distance of NGC 5204, can be included into the slit used in the
observation. In this case, the strength of the $\lambda1306$ emission line is
larger than that of those standard O/B stars, in which the surrounding H II
regions are only partially included.

This binary scenario is different from normal Galactic HMXBs, in which
accretion is due to stellar wind capture. 
Galactic HMXBs typically  have X-ray luminosities less than $10^{36}$ erg/s due
to the low efficiency of stellar wind accretion. 
In this case strong beaming effects would be required to boost the luminosity
above $10^{39}$ erg/s, which usually leads to small duty cycles. 
This is not the case for the ULX in NGC 5204, for which all 6 X-ray
observations in the past 10 years were above $10^{39}$ erg/s. 
In this binary scenario, matter accretes onto the primary by Roche Lobe
overflow, which is more efficient than stellar wind accretion. Some Galactic
LMXBs accrete by Roche Lobe overflow, and their X-ray luminosities can be as
high as $10^{39}$ erg/s. This will reduce the beaming factor required, if
required at all, and increase the duty cycle, thus helping to explain the X-ray
observations.

The period for such a system can be predicted given the mass and radius of the
secondary and the mass of the primary.
The mass and radius of the secondary can be determined empirically from its MK
spectral type, then the period will solely depend on the primary mass under the
Roche lobe overflow assumption.
For the B0 Ib secondary, we take a mass of 25 $M_\odot$ and a radius of 30
$R_\odot$ (these are values for B0 Iab taken from Schmidt-Kaler 1982).
The binary system would have a period of $\sim$204/290/295/284 hours for a
primary of mass 3/50/100/1000 $M_\odot$ (Liu et al.  2002). 
Long term monitoring of this object may lead to a mass estimate and help to
settle the debate whether the primary is a black hole of mass $10^3$--$10^4$
$M_\odot$ or a stellar mass black hole.

\acknowledgements

We are grateful for the service of HST and Chandra Data Archives. We would like
to thank the anonymous referee for his helpful comments, and Charles Cowley,
Ulisse Munari, Richard Gray, Joseph Cassinelli, Jimmy Irwin, Renato Dupke, and
Eric Miller for helpful discussions.  We gratefully acknowledge support for
this work from NASA under grants HST-GO-09073.

\clearpage

\begin{deluxetable}{lcccccccccc}
\tablecaption{The optical counterpart to the ULX in NGC 5204}
\tabletypesize{\tiny}
\tablehead{
\colhead{} & \multicolumn{2}{c}{ACIS-S3} &\colhead{}& \multicolumn{4}{c}{WF2} &\colhead{}& \multicolumn{2}{c}{HRC}\\
\cline{2-3}  \cline{5-8} \cline{10-11} \\
\colhead{Object} & \colhead{RA} & \colhead{DEC} & \colhead{}& \colhead{X} & \colhead{Y} &
\colhead{RA} & \colhead{DEC}& \colhead{}& \colhead{RA} & \colhead{DEC}\\
}

\startdata    

J132940.1+582454 & 13 29 40.19 $\pm 0\farcs24$ & +58 24 53.9 $\pm 0\farcs16$ & & 380.6 & 197.9 & 13 29 40.30 & 58 24 54.0  &  & 13 29 40.17 & +58 24 54.3\\
ULX & 13 29 38.61 $\pm 0\farcs02$ & +58 25 05.7 $\pm 0\farcs01$ & & 272.9 & 332.2 & 13 29 38.72 & 58 25 05.8 &  & 13 29 38.58 & +58 25 06.1\\

\enddata
\end{deluxetable}

\begin{deluxetable}{lccccccccc}
\tablecaption{Photometric measurements for U1, U2, HST-1, and J132940.1+582454}
\tabletypesize{\tiny}
\tablehead{
\colhead{} & \multicolumn{4}{c}{STMAG (mag)} &\colhead{}& \multicolumn{4}{c}{$F_\lambda$ (erg/s/cm$^2$/\AA)}\\
\cline{2-5}  \cline{7-10} \\
\colhead{Object} & \colhead{F220W} & \colhead{F435W} & \colhead{F606W} & \colhead{F814W} & \colhead{} & \colhead{F220W} & \colhead{F435W} & \colhead{F606W} & \colhead{F814W} \\
}

\startdata

U1 & 19.99 $\pm0.05$ & 21.92 $\pm0.05$ & 22.97 $\pm0.09$ & 24.03 $\pm0.08$ && 3.664e-17 & 6.194e-18 & 2.362e-18 & 8.872e-19 \\
U2  & 21.42 $\pm0.05$ & 23.67 $\pm0.05$ & 24.86 $\pm0.08$ & 25.72 $\pm0.21$ && 9.817e-18 & 1.236e-18 & 4.115e-19 & 1.872e-19 \\
HST-1 & 19.18 $\pm0.05$ & 20.41 $\pm0.05$ & 20.63 $\pm0.01$ & 21.18 $\pm0.01$ && 7.727e-17 & 2.489e-17 & 2.025e-17 & 1.226e-17 \\
J132940.1+582454 & 20.19 $\pm0.05$ & 21.77 $\pm0.05$ & 22.79 $\pm0.09$ & 23.81 $\pm0.08$ && 3.048e-17 & 7.112e-18 & 2.780e-18 & 1.086e-19 \\
\enddata
\tablecomments{
$F_\lambda$ is related to STMAG by $F_\lambda = 10^{-0.4\times(STMAG+21.1)}$.
}
\end{deluxetable}

\clearpage

\begin{figure}
\plottwo{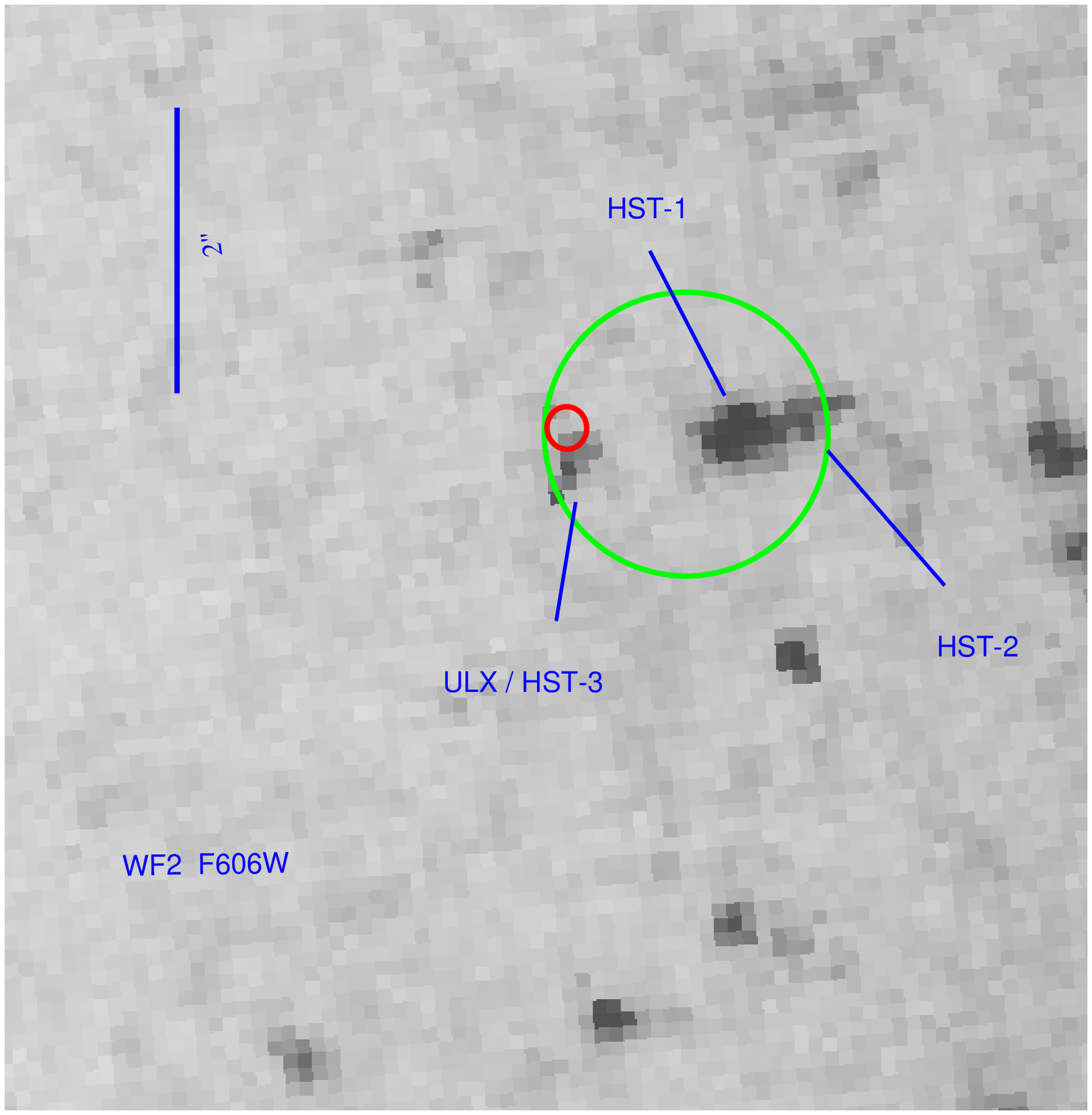}{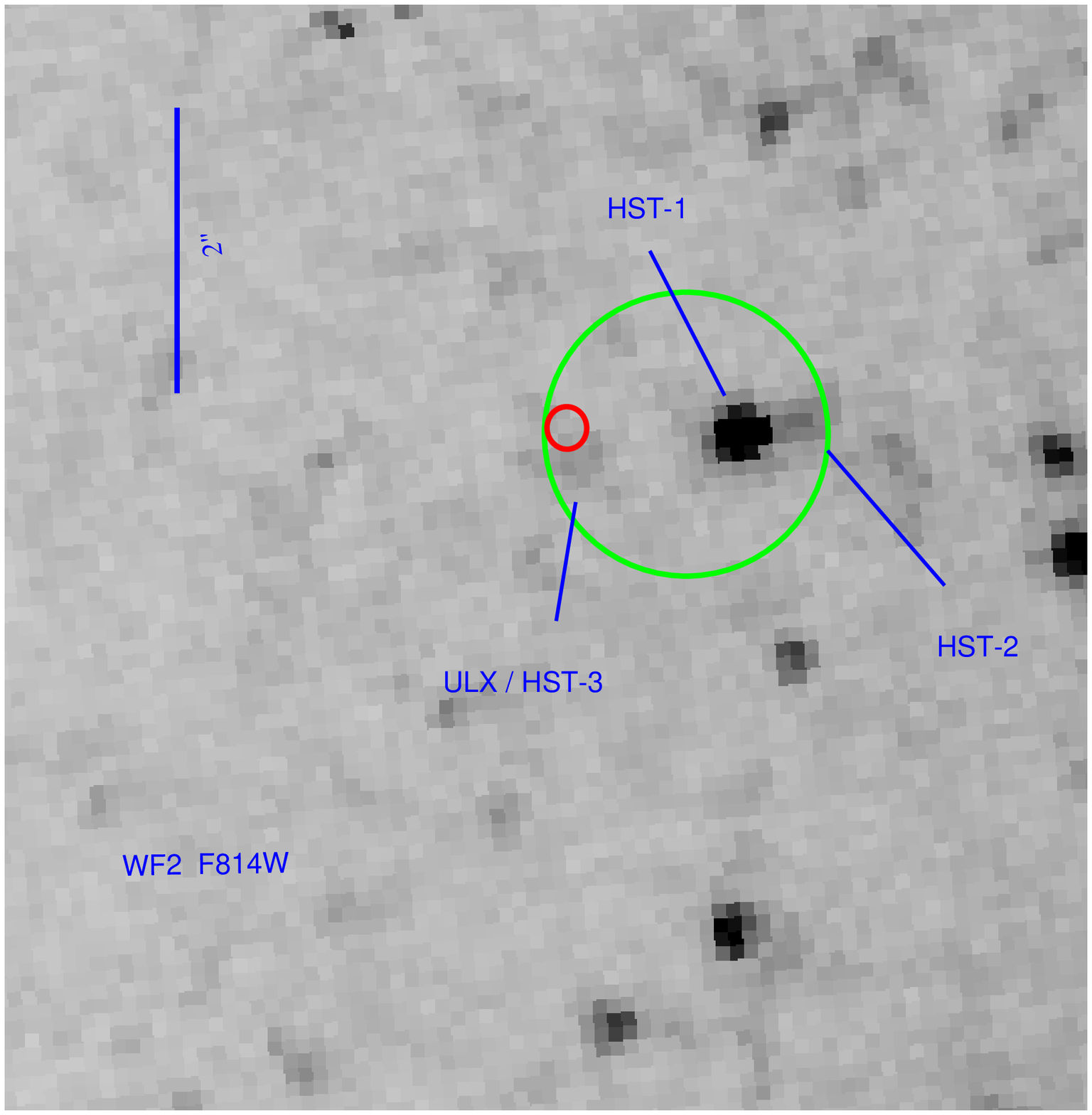}
\end{figure}
\begin{figure}
\plottwo{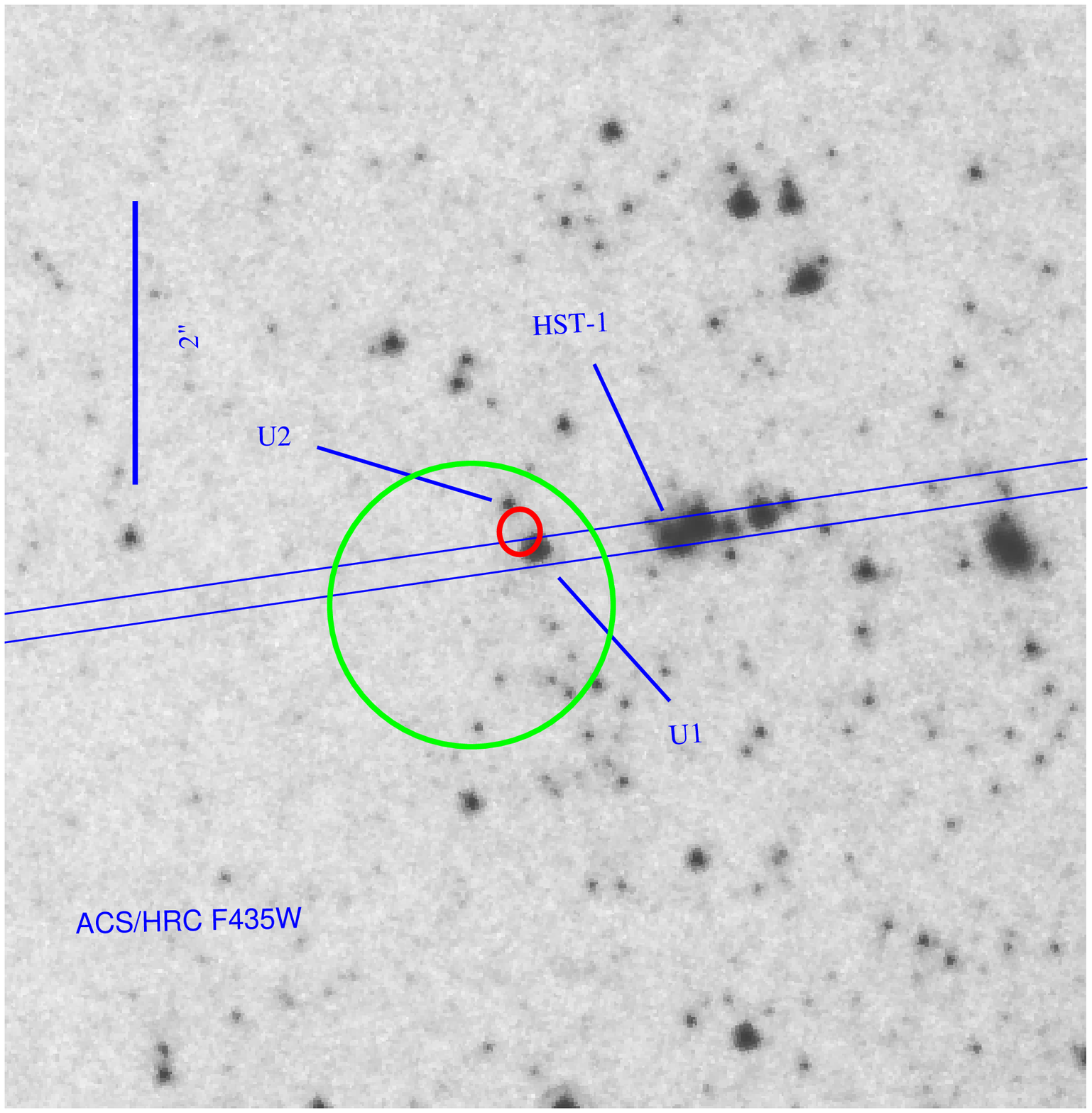}{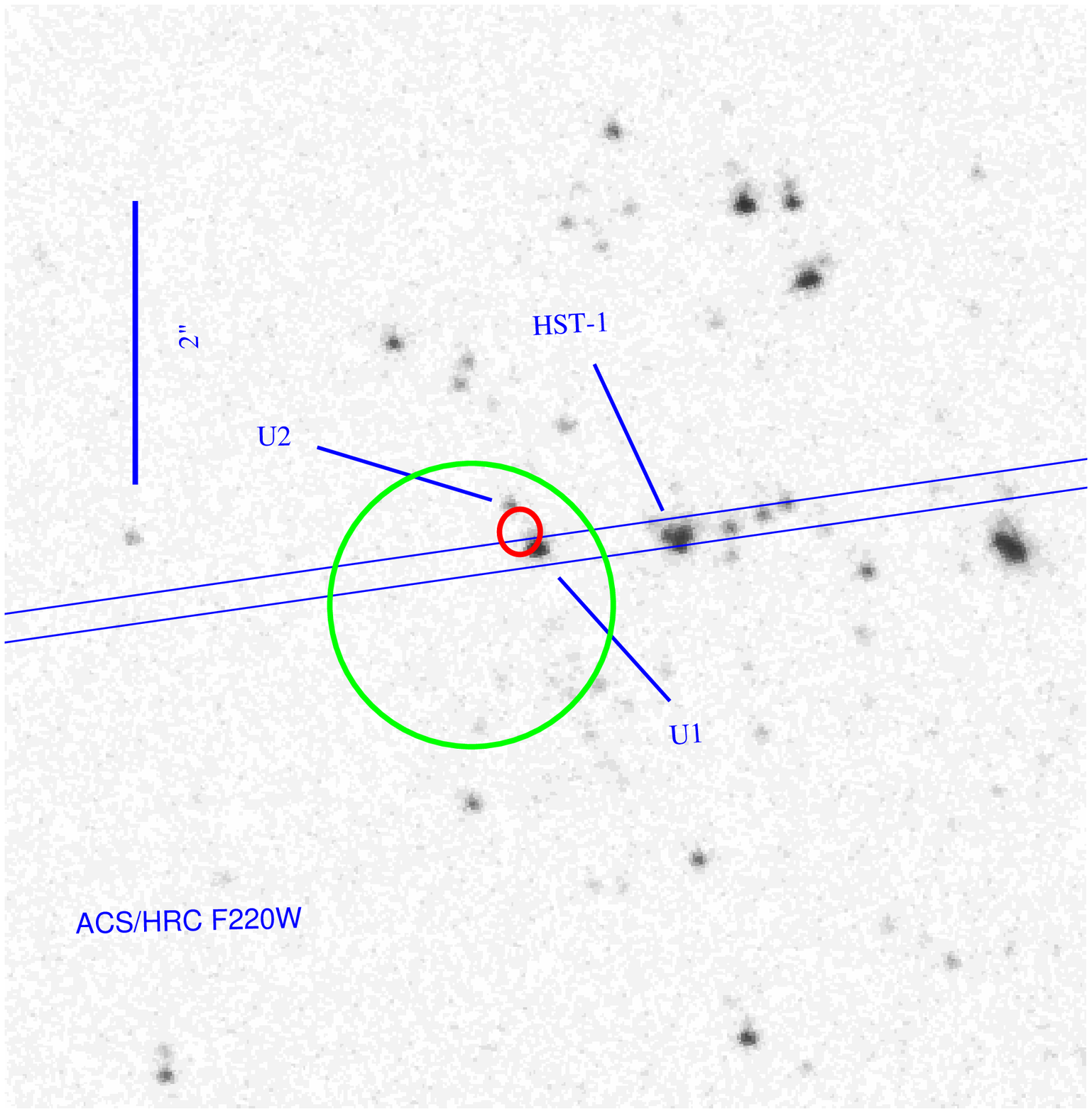}
\caption{The optical counterpart to the ULX in NGC 5204.  The small error
ellipses, shown on the WF2 images and the HRC images, is described in the text.
The $1^{\prime\prime}$ error circle in the WF2 image around the nominal Chandra
position was adopted by Goad et al.(2002); it includes different objects in the
HRC images due to the pointing offsets between WFPC2 and ACS. The slit position
for the STIS MAMA/FUV observation is shown in the HRC image.  }

\end{figure}

\begin{figure}
\plottwo{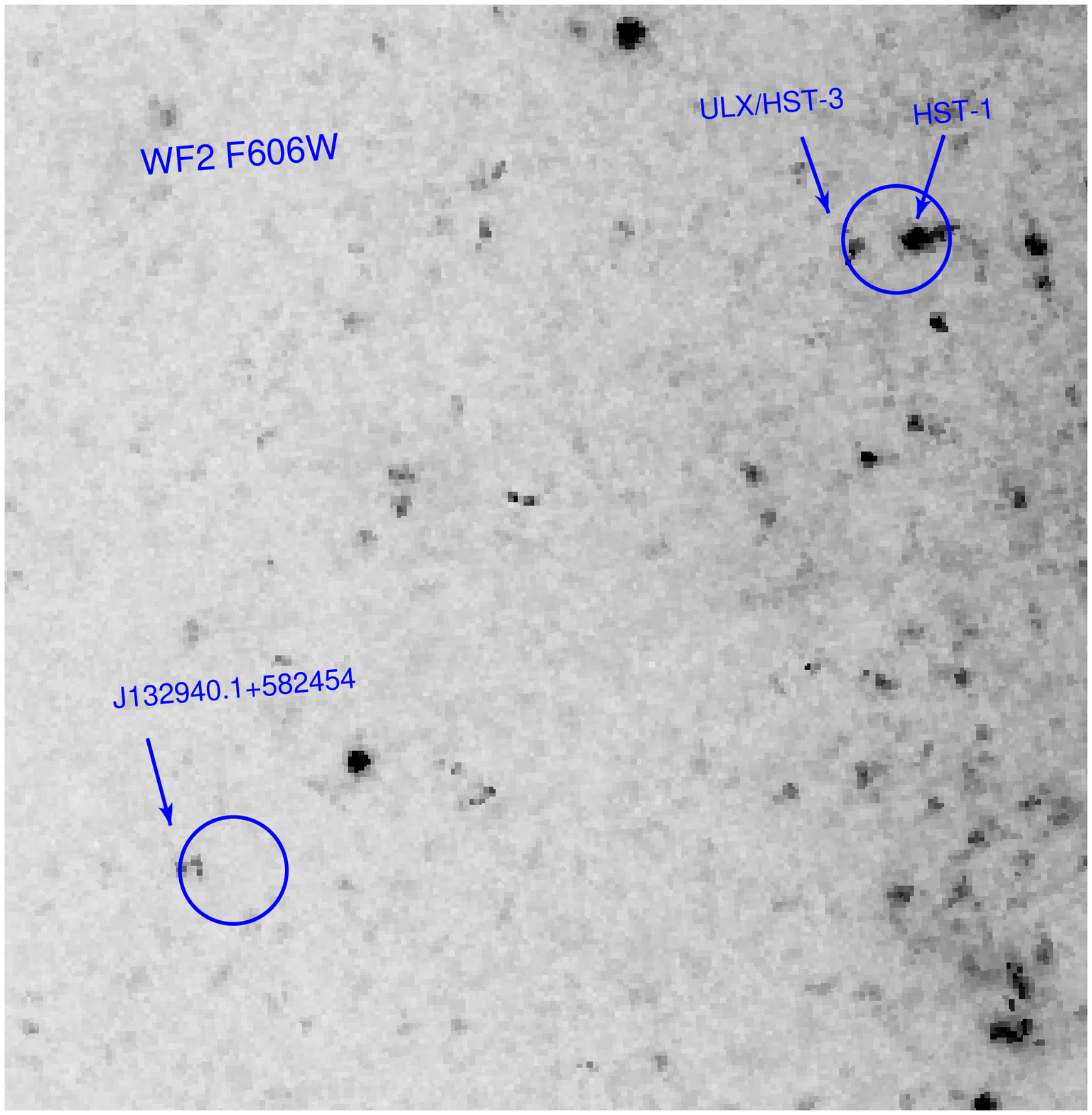}{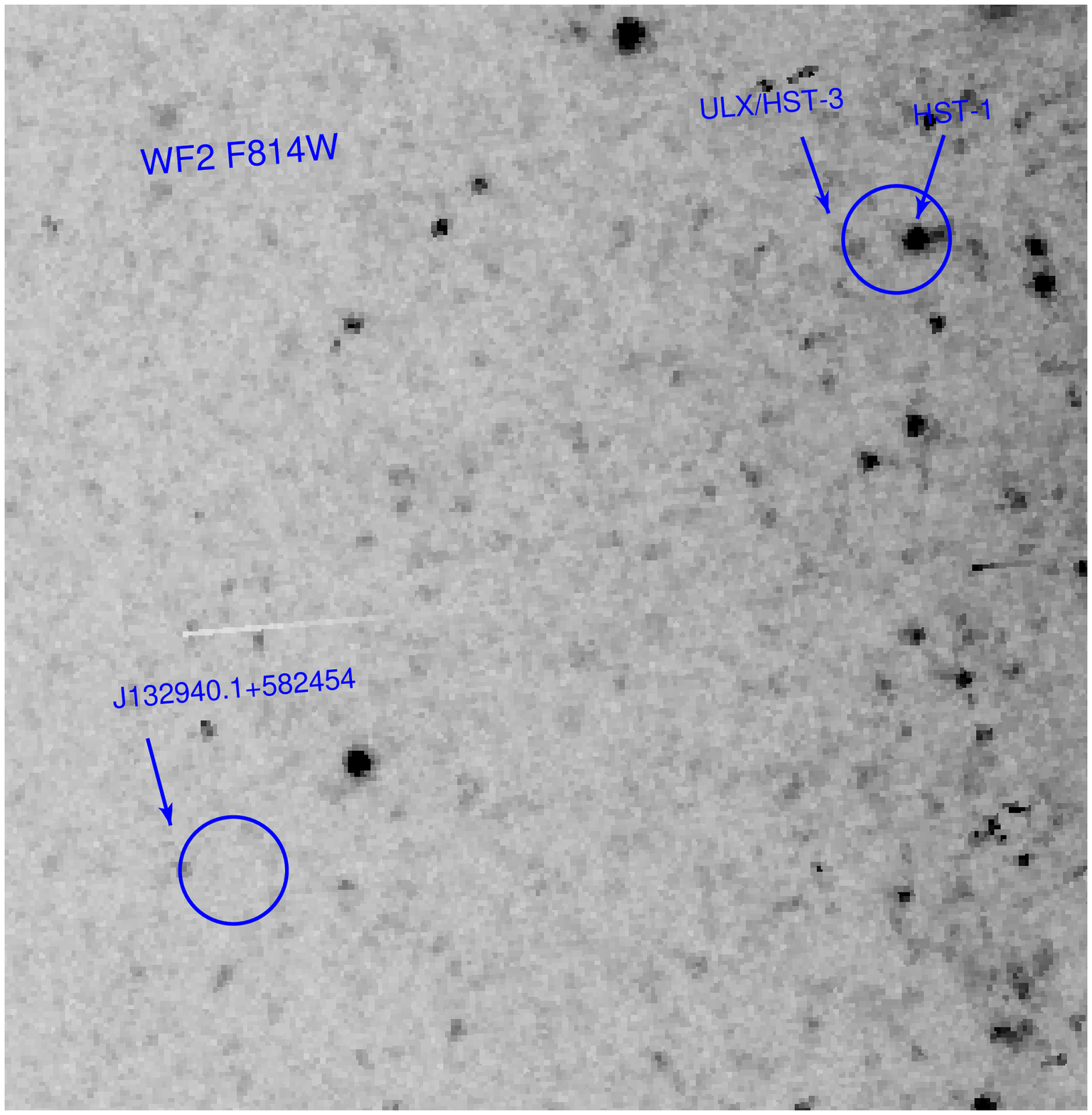}
\end{figure}
\begin{figure}
\plottwo{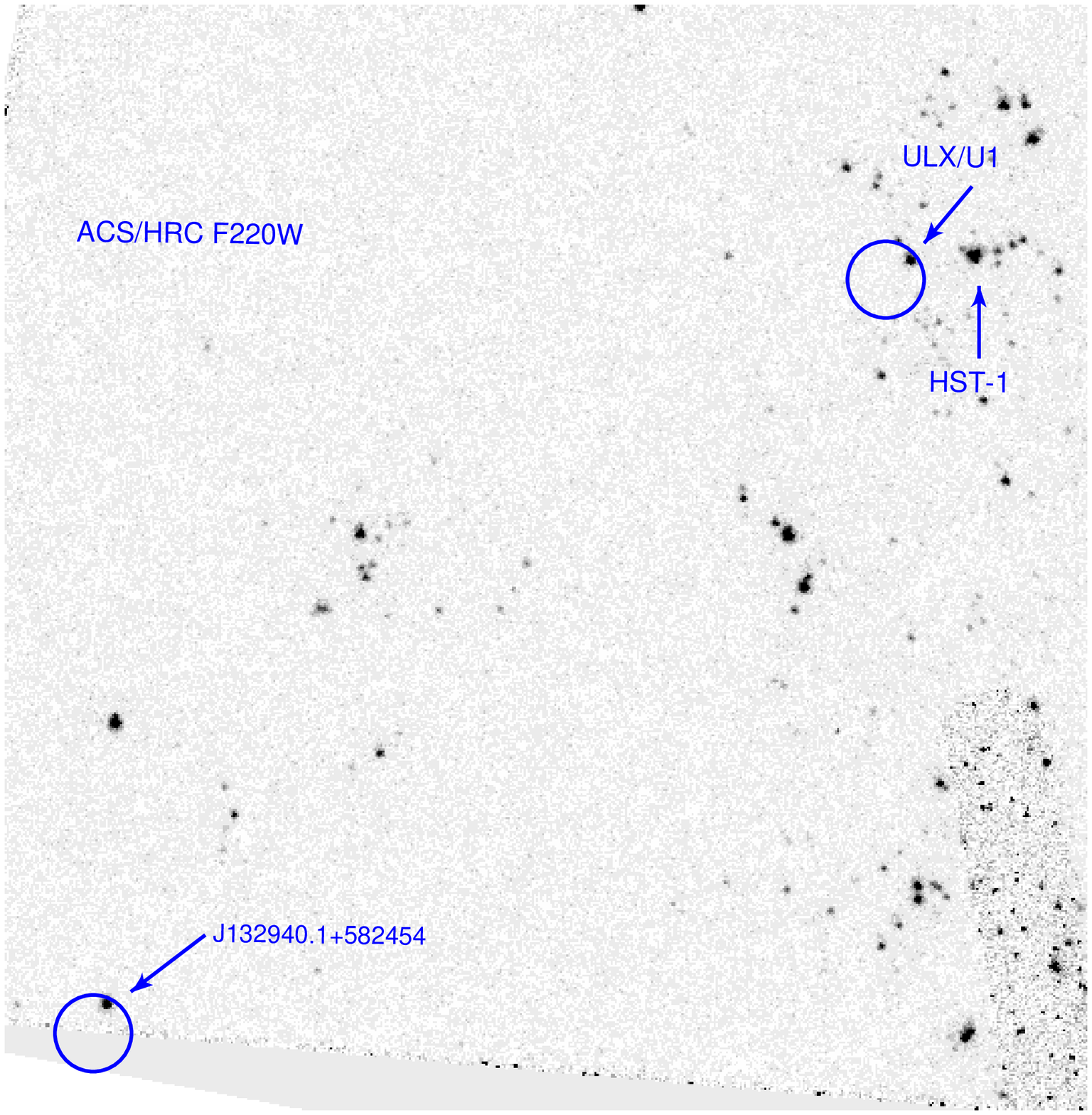}{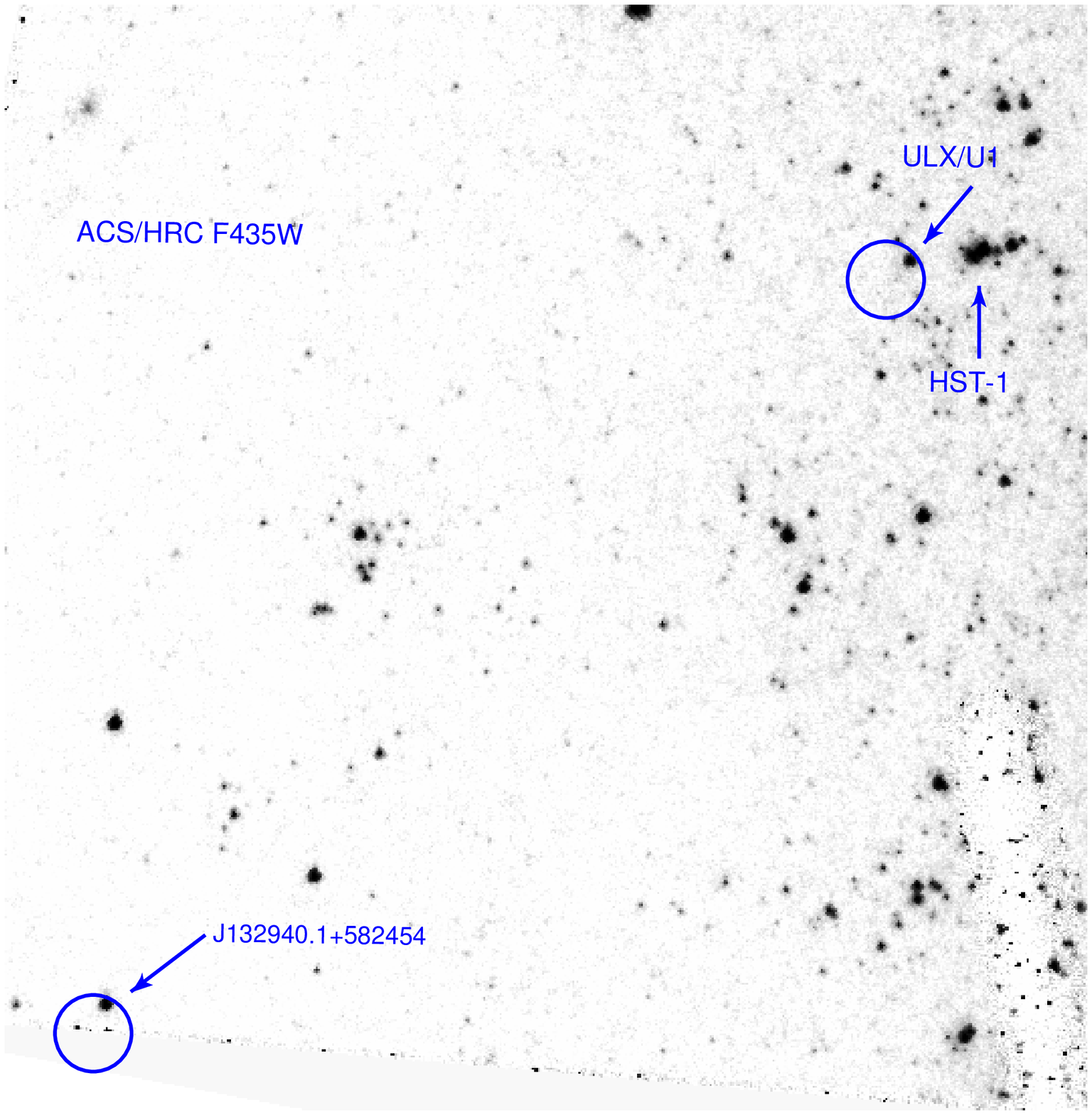}
\caption{The optical counterparts to J132940.1+582454 and the ULX.   
stands for J132940.1+582454.  The error circles on WF2 images have radii of
$1^{\prime\prime}$, while the error circles on the HRC images have radii of
$0\farcs6$.  }

\end{figure}

\begin{figure}
\plottwo{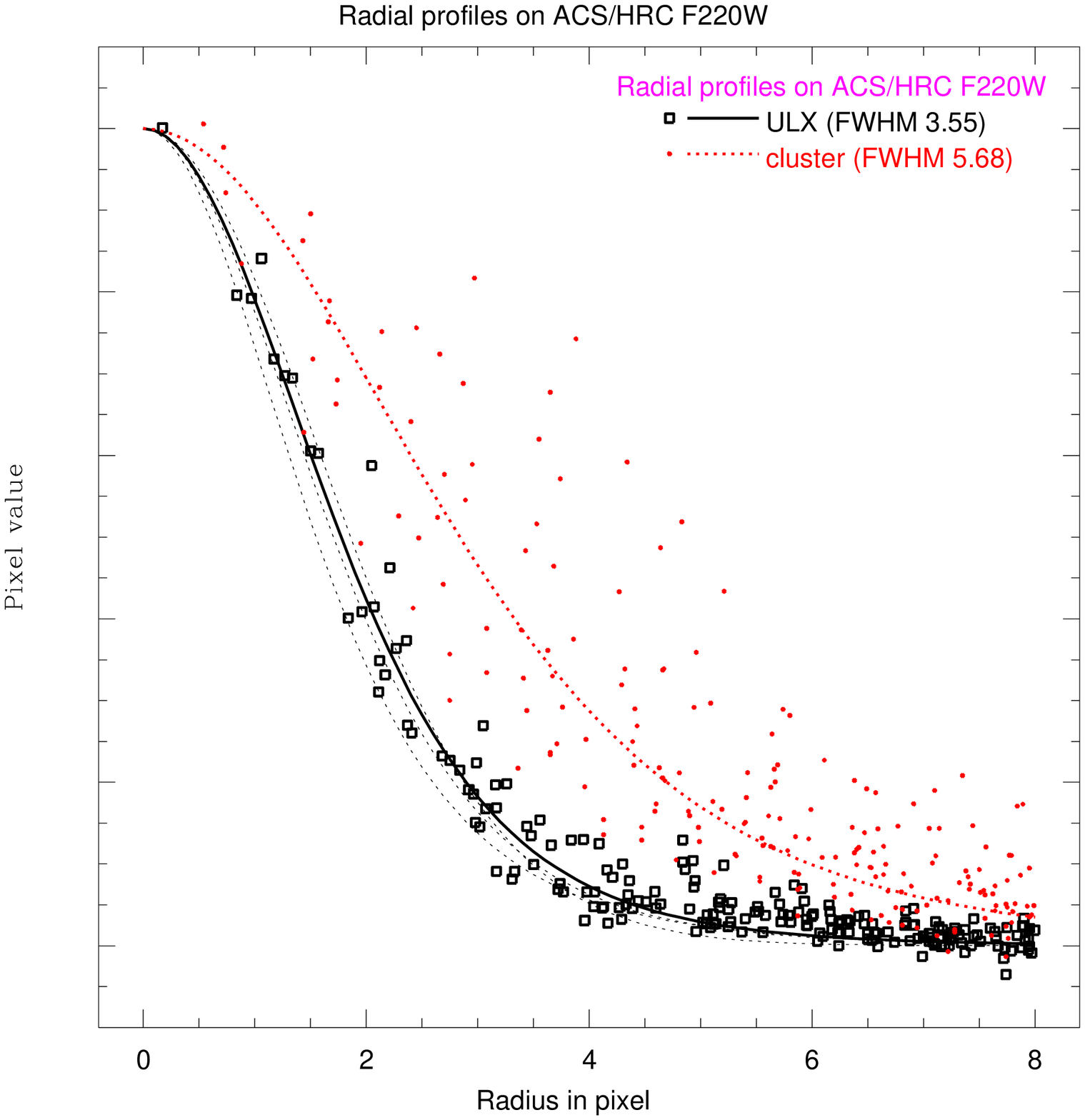}{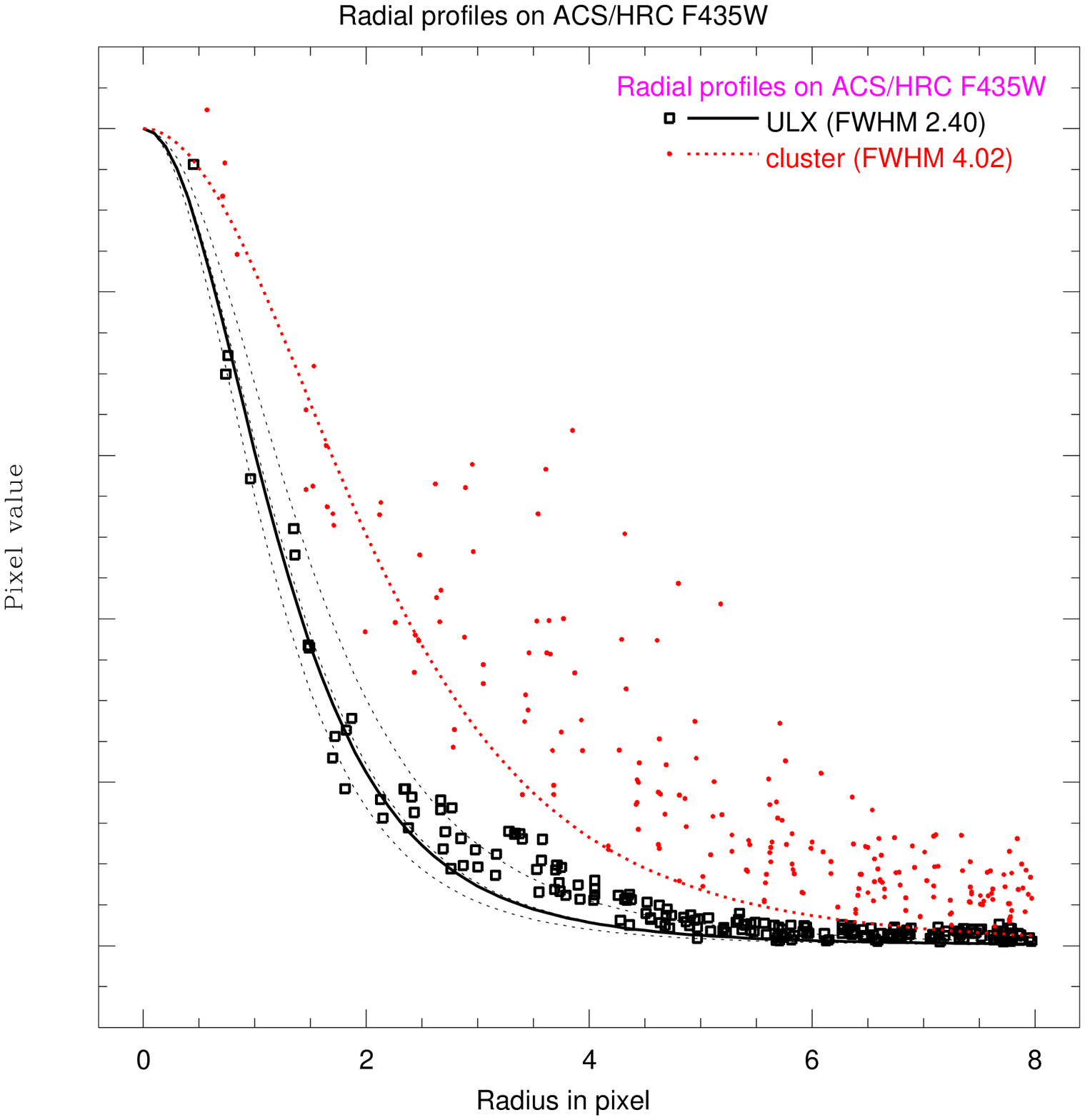}
\caption{The radial profiles of U1 in HRC F220W and F435W CCD images. The
squares and the thick solid line are the data and fit for the optical
counterpart U1. Fits for some typical stellar profiles are plotted in thin
dotted lines.  For comparison radial profiles for a nearby cluster 
are plotted in thick dotted lines and crosses.  In both images U1 is consistent
with being a point source.}

\end{figure}

\begin{figure}
\plotone{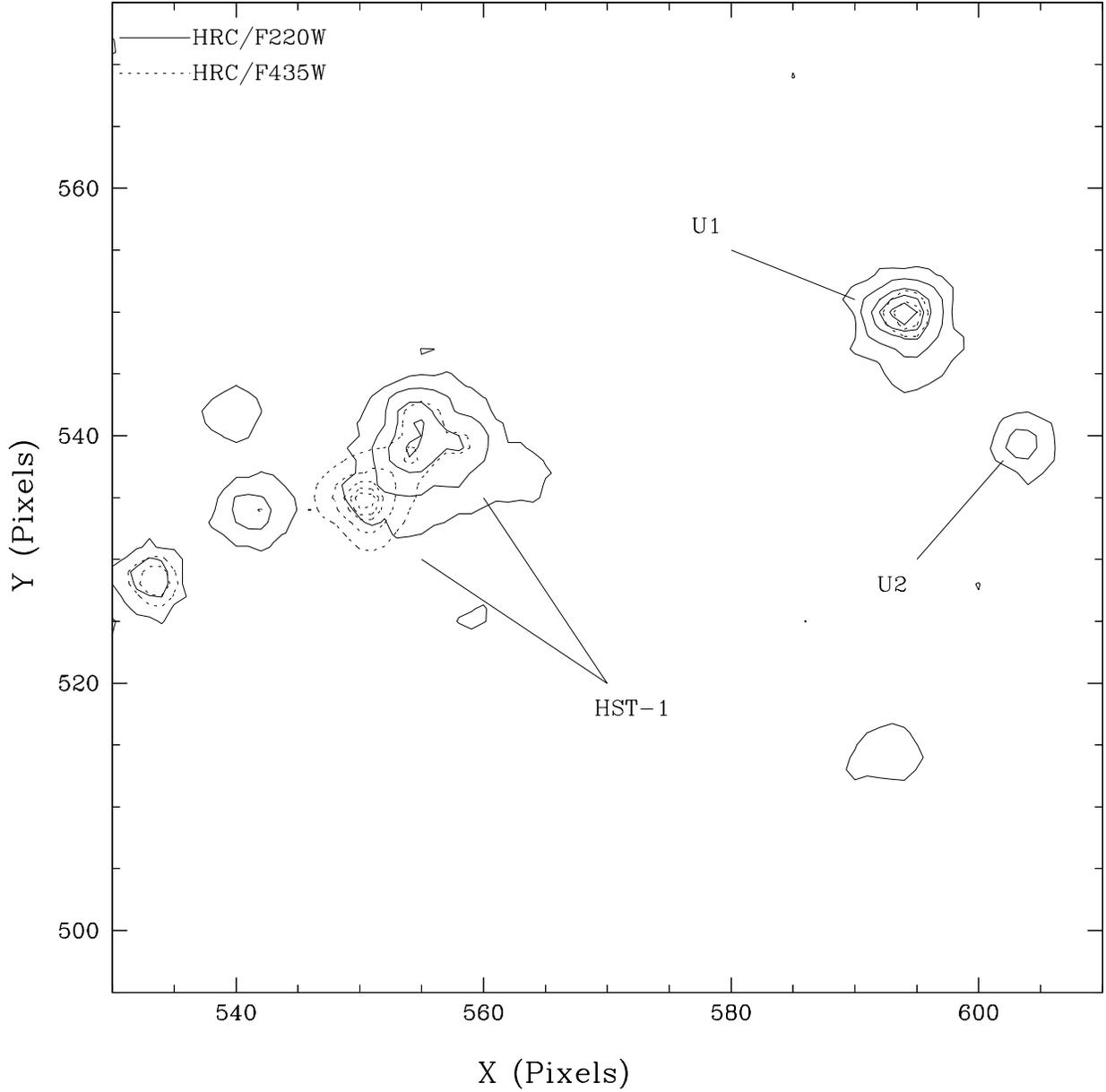}
\caption{Contour plots for the HRC F220W and F435W images. A 80 pixel
$\times$80 pixel region including U1, U2 and HST-1 is shown.  The optical
counterpart U1 has a compact point-like core.  HST-1, as one object in WF2
images, is now resolved into two components that are $\sim$7 pixels apart.}
\end{figure}

\begin{figure}
\plotone{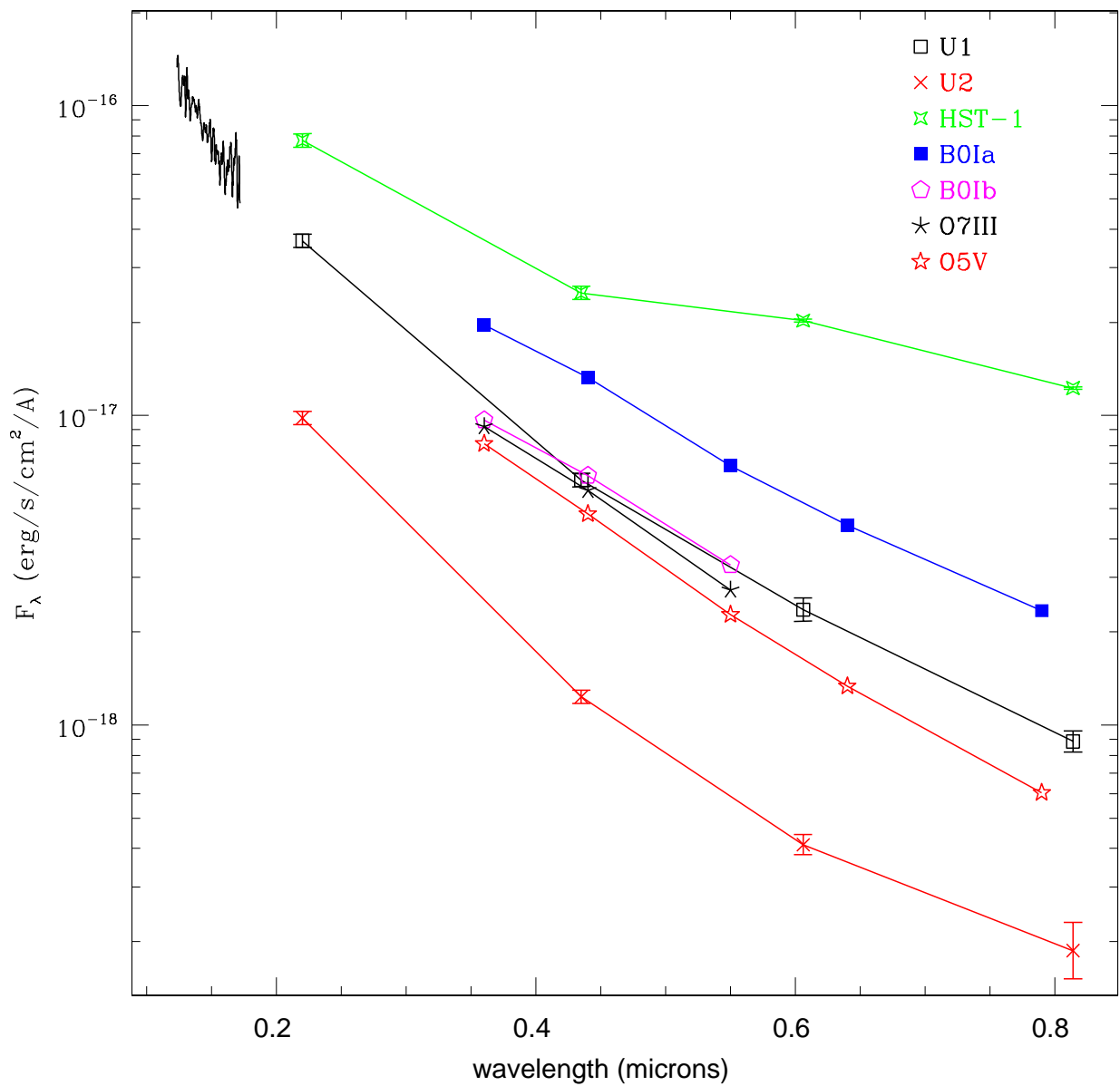}
\caption{Wide-band photometry and FUV spectrum for the optical counterpart to
the ULX in NGC 5204.  Fluxes at 220\AA, 435\AA, 606\AA, and 814\AA\ for U1, U2
and HST-1 are plotted.  For comparison are four standard stars placed with the
ULX in NGC 5204 of types O5 V, O7 III, B0 Ia and B0 Ib, to which extinction
inferred from X-ray absorption ( $n_H = 10^{21}$ cm$^{-2}$, $R_V$ = 3.1) is
applied.  The turn-up in the F606W and F814W fluxes of HST-1 is due to the red
component of this composite source.  }

\end{figure}

\begin{figure}
\plottwo{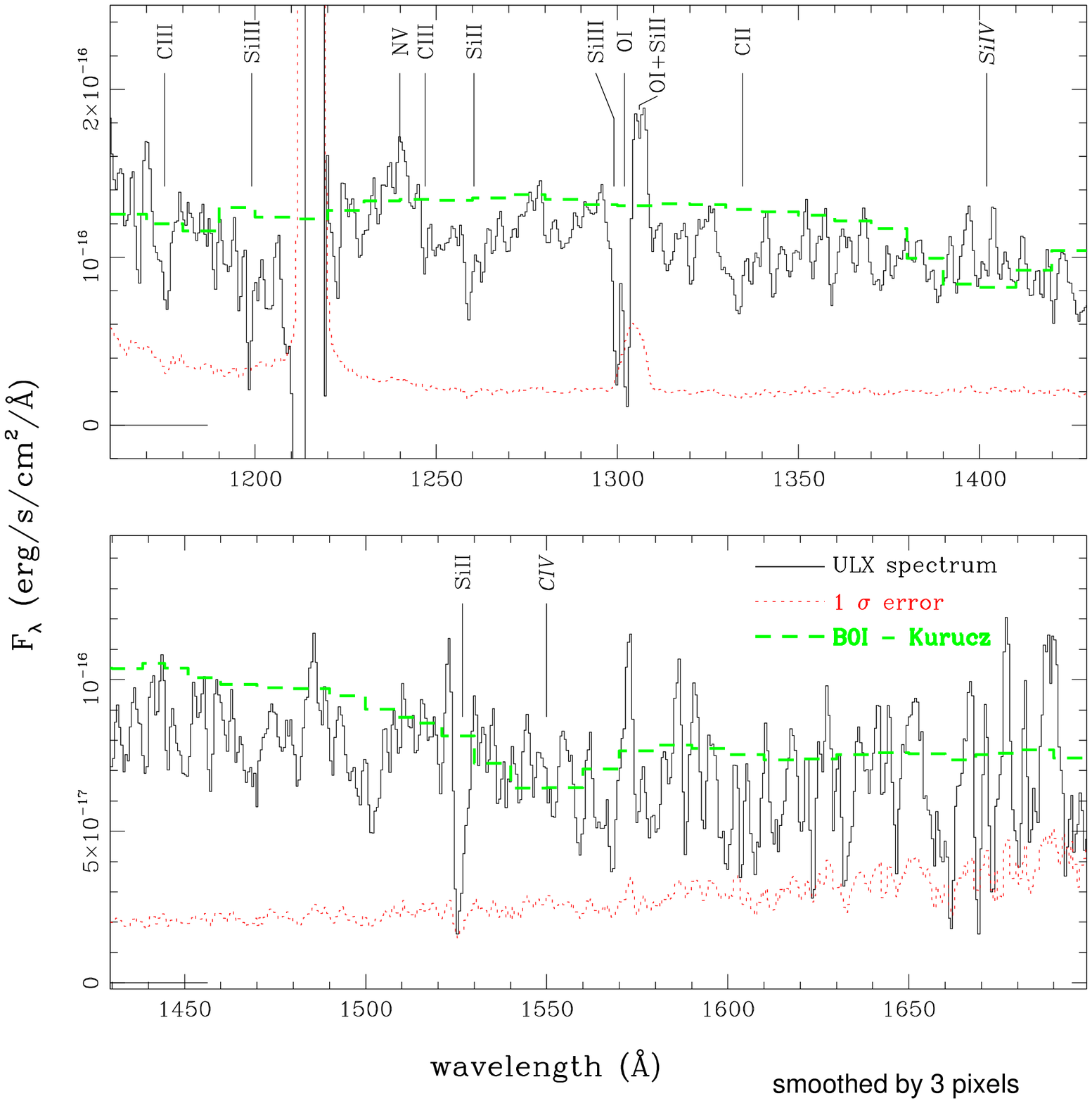}{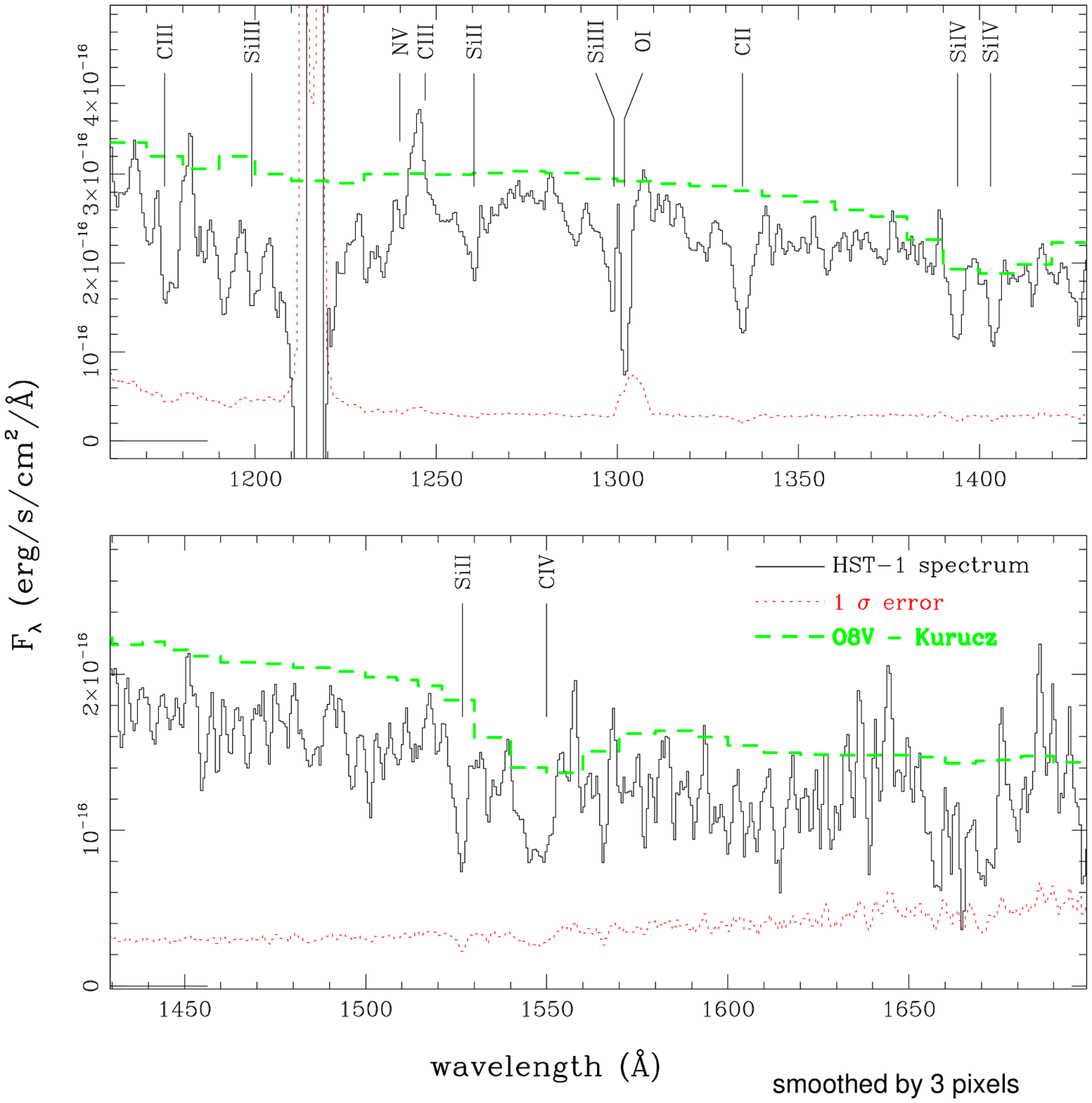}
\caption{HST STIS MAMA/FUV spectra of U1 (left) and HST-1 (right).  The
1$\sigma$ error is plotted in dotted line. Low resolution model spectra from
the Kurucz 1993 models are plotted in thick dashed lines for comparison.  The
O/B stellar features $\lambda1550$ C IV  and $\lambda1402$ Si IV missing from
U1 spectrum are labeled in italic.  }

\end{figure}
\end{document}